\documentclass[9pt,twocolumn,twoside]{pnas-new}
\usepackage{siunitx}
\usepackage{bm}
\usepackage[normalem]{ulem} 
\templatetype{pnasresearcharticle}

\definecolor{darkgreen}{rgb}{0,0.4,0.0}

\begin{document}
\pagestyle{empty}

\title{Active bacterial baths in droplets}

\author[a,1]{Cristian Villalobos-Concha}
\author[b,c,1]{Zhengyang Liu}
\author[a,d]{Gabriel Ramos}
\author[b,c]{Martyna Goral}
\author[b,f]{Anke Lindner}
\author[c]{Teresa López-León}
\author[b,f,2]{Eric Clément}
\author[a,2]{Rodrigo Soto}
\author[a,2]{María Luisa Cordero}

\affil[a]{Departamento de Física, FCFM, Universidad de Chile, Santiago, Chile.}
\affil[b]{Laboratoire PMMH, UMR 7636 CNRS-ESPCI-Sorbonne, PSL Research University, Sorbonne Université-Université de Paris, 7-9 quai Saint-Bernard, 75005 Paris, France.}
\affil[c]{Laboratoire Gulliver, UMR 7083 CNRS, ESPCI Paris, PSL Research University, 75005 Paris, France.}
\affil[d]{Laboratoire Softmat, Université de Toulouse, CNRS, UMR 5623, Université Toulouse III – Paul Sabatier, 31062, Toulouse, France.}
\affil[f]{Institut Universitaire de France, France.}

\leadauthor{Villalobos Concha}

\significancestatement{
Bacterial suspensions are a quintessential example of active matter, where each bacterium extracts energy from its environment to self-propel and, crucially for this study, to perturb the surrounding fluid. Given that bacterial suspensions often exist in confined environments, this article examines and characterizes the velocity fluctuations generated by bacteria confined within a droplet less than one millimeter in size. By tracking the motion of tracers inside the droplet, we demonstrate that the suspension behaves as a stochastic bath with a finite memory lasting a fraction of a second and an intensity that increases with bacterial density and the available space. This behavior contrasts sharply with thermal baths, which are memoryless and exhibit size-independent intensities.
}

\authorcontributions{Author contributions: A.L., T.L.-L., E.C., M.L.C., and R.S. designed the research; C.V.-C. and Z.L. conducted the research; C.V.-C., Z.L., E.C., M.L.C., and R.S. wrote the paper; all authors revised the final version.}
\authordeclaration{The authors declare no competing interest.}
\equalauthors{\textsuperscript{1}C.V.-C contributed equally to this work with Z. L.}
\correspondingauthor{\textsuperscript{2}To whom correspondence should be addressed.
E-mail: eric.clement@upmc.fr, rsoto@uchile.cl, mcordero@ing.uchile.cl}


\begin{abstract}
Suspensions of self-propelled objects represent a novel paradigm in colloidal science. In such “active baths”, traditional concepts, such as Brownian motion, fluctuation-dissipation relations, and work extraction from heat reservoirs, must be extended beyond the conventional framework of thermal baths. Unlike thermal baths, which are characterized by a single parameter, the temperature, the fundamental descriptors of an active bath remain elusive, especially in confined environments. In this study, buoyant passive tracers are employed as generalized probes to investigate an active bath comprising motile bacteria confined within a droplet. We demonstrate that momentum transfer from the bath to the tracer can be effectively described as colored noise, characterized by temporal memory and an enhanced effective diffusivity significantly larger compared to thermal Brownian motion values. Using a stochastic analytical framework, we extract the temporal memory and diffusivity parameters that define such an active bath. Notably, the diffusivity scales linearly with bacterial concentration, modulated by a factor representing the role of confinement, expressed as the ratio of the confining radius to the probe radius. This finding, while still awaiting a complete theoretical explanation, offers new insights into the transport properties of confined active baths and paves the way for a deeper understanding of active emulsions driven by confined active matter.
\end{abstract}

\dates{This manuscript was compiled on \today}
\doi{\url{www.pnas.org/cgi/doi/10.1073/pnas.XXXXXXXXXX}}

\maketitle
\thispagestyle{firststyle}
\ifthenelse{\boolean{shortarticle}}{\ifthenelse{\boolean{singlecolumn}}{\abscontentformatted}{\abscontent}}{}

\section*{Introduction}

At the beginning of the last century, one of the great achievements of Albert Einstein was to establish a direct connection between the erratic motion of a colloidal particle and the thermal energy contained in molecular environments at thermodynamic equilibrium~\cite{Einstein_1905}. This was the dawn of a deep understanding of the notion of thermal baths that relates thermal energy to the level of fluctuations expressed by all degrees of freedom associated with a return to thermodynamic equilibrium. Quantitatively, this effect is captured by a set of generic fluctuation-dissipation relations. Almost one century later, in a seminal paper, Wu and Libchaber~\cite{wuParticleDiffusionQuasiTwoDimensional2000} have shown that, in the case of passive tracers moving in a suspension of motile bacteria confined in a thin film, the relation between tracer diffusivity and bacteria activity goes beyond the Stoke-Einstein's fluctuation-dissipation relation. Further works identified this diffusion enhancement as stemming from a fundamental time-irreversibility borne by tracer/swimmer interactions~\cite{Dunkel2010, Thiffeault2010, Mino2013, Pushkin_2014, Morozov_2014, Kasyap2014, Burkholder_2017, Mathijssen_2018, Lagarde_2020}, hence revealing novel and exciting properties attached to the concept of ``active baths'' that involves motile swimmers either of biological or artificial origin~\cite{Kim_2004, Leptos_2009, Mino2011, Ginot_2015}. The presence of microscopic sources of energy and momentum transfer has been related to a great wealth of novel and counter-intuitive phenomena such as spontaneous bacteria re-concentration~\cite{Galajda2007}, work extraction via ratcheted wheels~\cite{Sokolov2010, DiLeonardo2010}, large scale collective motion~\cite{dombrowski2004self}, active colloidal clustering~\cite{Bouvard2023}, and the emergence of new structural phases~\cite{Petroff2015}. From a fundamental point of view, this also led to reconsidering the thermodynamic principles for out-of-equilibrium situations, hence revisiting fluctuation theorems and the generic efficiency of molecular machines~\cite{Seifert_2012, OByrne_2022}. One of the most crucial questions that emerged was related to the proper role of boundaries, as they usually play a key role in the dynamics of the system, radically different from what is expected for standard thermal matter~\cite{Takatori_2014, Smallenburg_2015, Nikola_2016}. For such active baths, specific boundary geometries are able to control the motion of active agents, to rectify their trajectories~\cite{Galajda2007}, lead to specific surface accumulations~\cite{Berke2008} and contribute to detailed-balance violations that drive the spontaneous emergence of macroscopic fluxes as well as mixing properties, without counter-parts in classical thermal equilibrium. Along those lines, monitoring the motion of passive particles in bacterial suspensions serves as a valuable tool to probe the active suspension properties. By analyzing the statistical properties of the particle trajectories, such as mean squared displacement, velocity auto-correlation, and probability distribution functions, one can extract quantitative information on bacterial activity, on the effective viscosity of the suspension, and thus reveal the nature of the active noise experienced by the probe particles~\cite{maggiGeneralizedEnergyEquipartition2014, maggi2017memory, chaki2018entropy, yeActiveNoiseExperienced2020, dabelow2019irreversibility}.

In this report, we question the role of spherical confinement on an active bacterial bath. This situation is important not only from a conceptual viewpoint but also from a practical perspective, as droplet encapsulation of bacteria is commonly used in many emergent bio-technologies as high-throughput bioreactors~\cite{Kaminski2016, Pryszlak2021, Postek2022}. In climate science, it is also known that bacteria populations are naturally present in cloud droplets and can crucially participate in the triggering of precipitations~\cite{Christner2008}. Moreover, this situation can be more generally viewed as a simplified toy model featuring the important ``crowded cell problem'' where colloidal probes \cite{Wilhelm_2010}, various organelles, or oil droplets (for adipocytes) are driven or stirred by the out-of-equilibrium cellular activity~\cite{Norregaard_2017, Alfano_2024}.

To address all these questions, we propose an extensive study where positively or negatively buoyant spherical objects, namely oil droplets or solid particles, are encapsulated in an inverse emulsion of swimming bacteria. In our experimental model, confinement can be modified by two complementary means: first, by controlling the size of the confining drop, thus varying the space where the active agents and the passive tracer are located, and second, by varying the size of the tracer, which also affects the available space for the active environment. We explore the passive tracer kinematics for a large range of radii (confining droplet and tracer). The bacteria concentration is varied over several orders of magnitude. To monitor the kinematics of the encapsulated object, complementary techniques are used, either by 2D tracking on perpendicular projection planes or by following the tracer directly in 3D, using a Lagrangian tracking technique.

\section*{Results}

\subsection*{Definition of the experimental system}
\begin{figure*}[t!]
    \centering
    \includegraphics[width=\linewidth]{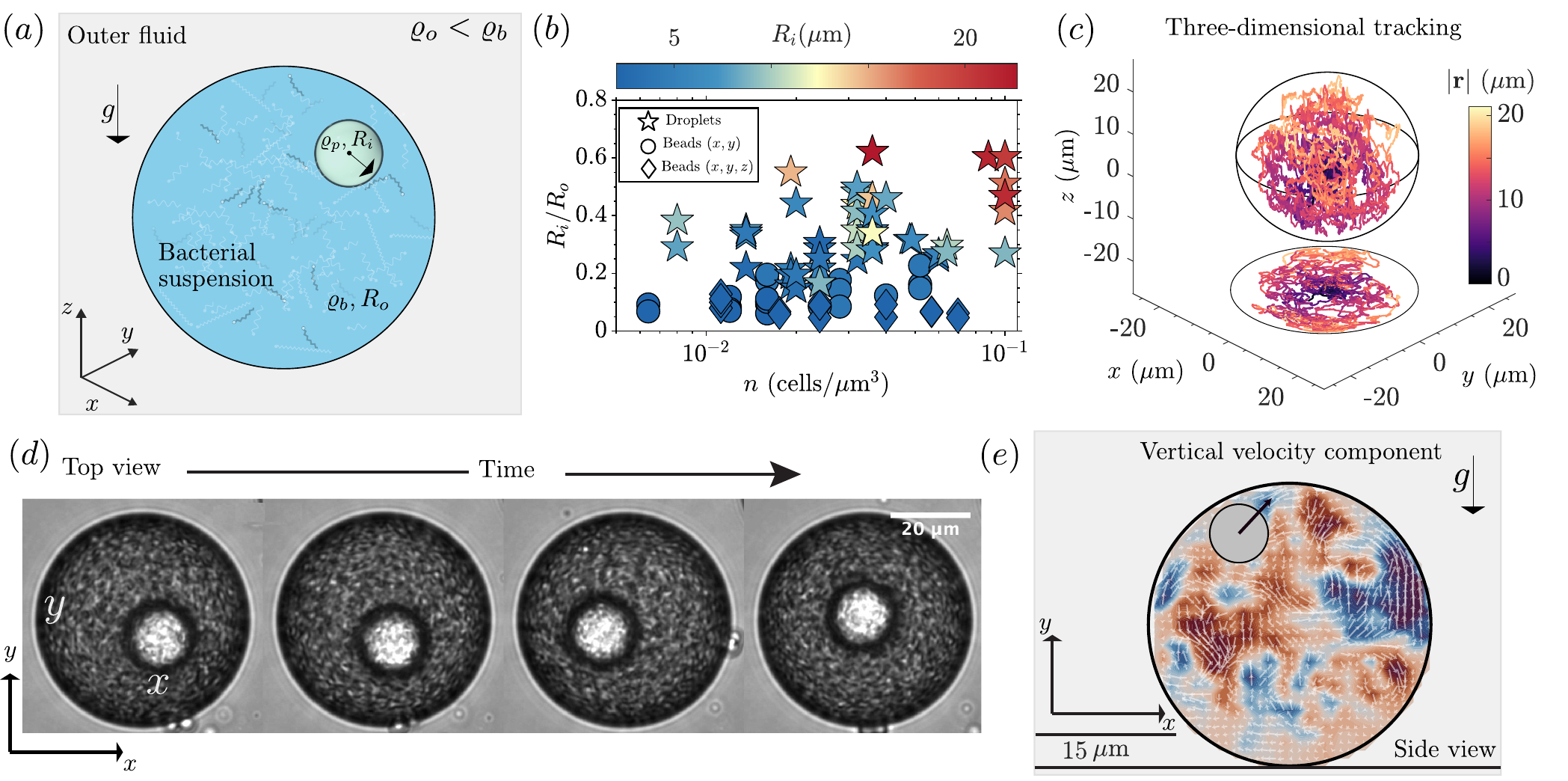}
    \caption{(a) A schematic of a passive particle of radius $R_i$ and density $\varrho_p$ confined to a spherical bacterial droplet of radius $R_o$ and density $\varrho_b$. The droplet is immersed in a fluid of density $\varrho_o$.
    (b) Parameter space of the experimental realizations. Markers of different shapes indicate the different setups.
    (c) Three-dimensional trajectory of a melamine particle of radius $R_i = \SI{2}{\micro\meter}$ confined within a droplet of radius $R_o = \SI{23}{\micro\meter}$ with a bacterial suspension of cell density $n = \SI{2.4e-2}{cells/\micro\meter^3}$. The colorbar indicates the instantaneous radial position $|\mathbf{r}| = \sqrt{x^2 + y^2 + z^2}$.
    (d) A sequence of images showing the $(x,y)$ view of the double emulsion. The parameters are $R_o = \SI{31}{\micro\meter}, R_i = \SI{6.5}{\micro\meter}$, and $n = \SI{6.4e-2}{cells/\micro\meter^3}$. See also \textit{SI Appendix}, Movie S1.
    (e) Flow field in $(x,z)$ view of the double emulsion. The parameters are $R_o = \SI{31}{\micro\meter}, R_i = \SI{6.5}{\micro\meter}$, and $n = \SI{6.4e-2}{cells/\micro\meter^3}$.}
    \label{fig:fig_1}
\end{figure*}

The inverse emulsion of oil and motile bacterial suspension is produced by agitating a mixture of oil and an aqueous bacterial suspension. The resulting bacterial droplets are heavier than the suspending oil and thus sediment at the bottom of the observation chamber. The passive particle encapsulated within the spherical droplet can either be an oil droplet of negative buoyancy, spontaneously formed during the emulsification, or a solid particle of positive buoyancy, originally suspended in the bacterial suspension at low number density. In the absence of bacteria, this would cause, respectively, flotation at the top or sedimentation at the bottom. In all experiments, the tracers, either solid beads or oil droplets, indistinctly called particles, are spheres of radius $R_i$, and their density is denoted $\varrho_p$ (Fig.~\ref{fig:fig_1}(a)). The different values that $\varrho_p$ can take, $\varrho_p > \varrho_b$ for beads or $\varrho_p < \varrho_b$ for oil droplets, where $\varrho_b$ is the density of the bacterial suspension, are considered during analysis in the definition of the particle's buoyant mass $\Delta M = 4\pi R_i^3 \Delta \varrho /3$, with $\Delta \varrho = |\varrho_p - \varrho_b|$, while the different direction of sedimentation, in either case, does not change the dynamical behavior (besides a simple reflection of the vertical axis). A systematic characterization of the system is performed by using different particle sizes $R_i \in [2, 22]$~\si{\micro\meter}, confinement ratios $R_i/R_o \in [0.1, 0.6]$, and bacterial concentration $n \in [8 \times 10^{-4}, 1.2 \times 10^{-1}]$~\si{cell/\micro\meter^3}, as summarized in Fig.~\ref{fig:fig_1}(b). The degree of confinement of the particle is characterized either by the effective available space $R = R_o - R_i$ or by its dimensionless form $R/R_i$.

In Fig.~\ref{fig:fig_1}(c), we display the dynamics of a fluorescent passive bead agitated by the bacteria. To this end, we used an in-house 3D Lagrangian tracking technique~\cite{Darnige2017} that allows a full $(x,y,z)$ monitoring (for the definition of the coordinate system, see Fig.~\ref{fig:fig_1}). The tracks reveal complex three-dimensional kinematics. The region explored by the bead is not restricted to dwell around the bottom equilibrium position and, indeed, expands to a much larger volume within the droplet. In many situations, we observe that the beads detach from the inner droplet surface and perform long-time excursions into the central part of the drop. The outer droplet confinement restricts the bead movement and brings an upper limit for the inner probe displacement.

The 3D tracking data is complemented with 2D tracking in the $(x,y)$ plane obtained with bright field imaging. In Fig.~\ref{fig:fig_1}(d), we show an example of four successive inner oil droplet positions in the $(x,y)$ plane. In this case, the inner droplet remains essentially confined to the outer droplet wall and the tracking could be performed over long times. Observations in the $(x,z)$ plane are also possible by using another in-house experimental set-up that can tilt the microscope horizontally~\cite{Billon_2023}. In Fig.~\ref{fig:fig_1}(e), we display the track of an oil droplet monitored using this last technique. Vertical flows measured by a PIV technique can be obtained, and for example, large-scale flows were found to consistently occur when the probe particle leaves the surface (\textit{SI Appendix}, Fig. S1). Those measurements reveal the full complexity of the problem that inherently associates hydrodynamics, bacterial activity, and confinement. Indeed, a full bottom-up theoretical investigation would require highly involved hydrodynamic techniques, as recently exemplified by the theoretical study of magnetotactic bacteria (MTB) encapsulated within a droplet~\cite{Thery_2024}, that explain the self-assembly of MTB into a rotary motor under a magnetic field~\cite{Vincenti2019}.

\subsection*{Planar mean square displacement}
For a full 3D trajectory $(x(t),y(t),z(t))$ of the encapsulated probe, the exploration process can be characterized by the mean square displacement (MSD) computed for each direction and defined as $\langle \Delta r_i^2(t) \rangle = \langle [r_i(t + t_0) - r_i(t_0)]^2 \rangle$, averaged over $t_0$, with $i = x, y, z$. Due to the axial symmetry, the $x$ and $y$ components are equivalent and, therefore, we present their sum $\langle \Delta{\bm \rho}^2(t) \rangle$, which we refer to as the planar MSD, where $\bm{\rho} = x\hat{\mathbf{x}} + y\hat{\mathbf{y}}$.

\begin{figure}[t!]
    \centering
    \includegraphics[width=0.8\linewidth]{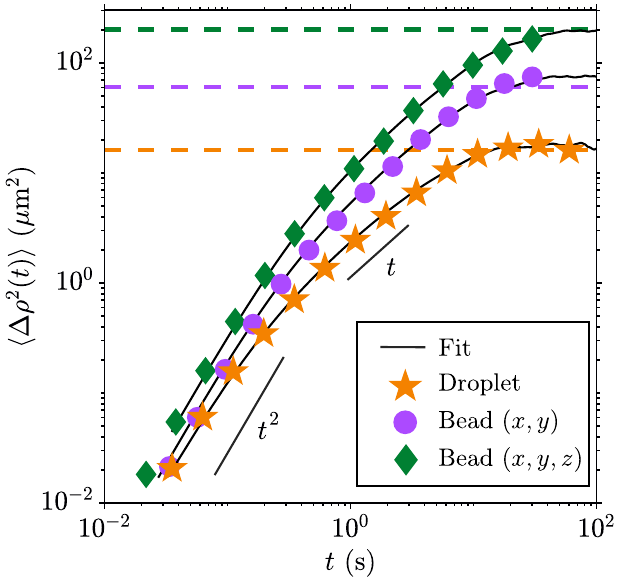}
    \caption{Typical planar MSD curves for the planar coordinate $\bm \rho$. The symbols represent the experimental data and the lines denote the best fitted MSD. The experimental parameters are: beads $(x,y,z)$, $R_o = \SI{23}{\micro\meter}$, $R_i = \SI{2}{\micro\meter}$, $n = \SI{1.1e{-2}}{cells/\micro\meter^3}$; beads $(x,y)$, $R_o = \SI{15}{\micro\meter}$, $R_i = \SI{2}{\micro\meter}$, $n = \SI{4e{-2}}{cells/\micro\meter^3}$; and double emulsion $(x,y)$, $R_o = \SI{35}{\micro\meter}$, $R_i =\SI{10}{\micro\meter}$, $n = \SI{6.4e{-2}}{cells/\micro\meter^3}$. The dashed horizontal lines represent twice the variance of $\bm \rho$ for each case.}
    \label{fig:fig_2}
\end{figure}

Although the $(x,y,z)$ and $(x,z)$ tracking can in principle provide a full characterization of the probe motion, the optical complexity of the setups limits the quality of the tracking and induces relatively large uncertainties in the vertical $z$ positions. Consequently, to characterize the particle motion, we will make use of the planar MSD, which can be accurately obtained for all tracking setups. Figure~\ref{fig:fig_2} presents three typical planar MSD curves obtained for three different experimental conditions. Distinct regimes can be observed. At short time scales, the particles exhibit a ballistic or an apparent super-diffusive behavior, characterized by a power-law dependence of the MSD on time with an exponent $\alpha \approx 1.9 \pm 0.1$. The fact that this exponent is slightly smaller than 2 is a known artifact arising from finite time sampling in experiments and detection noise in the particle position~\cite{martin2002apparent}. Indeed, a true ballistic regime at short times is expected to arise from the coherent active forces exerted by the swimming bacteria on the particle, which cause it to move in a correlated manner over a short time scale. After this ballistic regime, a diffusive regime can be observed in some cases, although it does not always appear. At longer times, the MSDs are bound to saturate as a consequence of the confinement imposed by the droplet boundary on the particle's motion. In a confined system, the saturation value of the MSD in any coordinate is equal to twice the variance of the particle's position in that coordinate (\textit{SI Appendix}, SI text), a relationship indicated by the dashed lines in Fig.~\ref{fig:fig_2}(a), confirming that experimental runs are sufficiently long to ensure the system has reached the ergodic limit.

The planar MSD contains, in principle, all relevant statistical information suited to characterize the bacterial bath acting on the particle. However, such a characterization requires an underlying model describing the particle dynamics, including a description of the action of the bacterial suspension on the tracer in the presence of the geometrical confinement. For the bacterial fluid, as explained in the next section, we opt for a model of a noisy bath with a local temporal correlation. For the confinement, one may be tempted to use a harmonic potential. Indeed, an analytical solution for the planar MSD of a passive particle in a harmonic potential subjected to a noisy bath with temporal memory was proposed earlier~\cite{maggiGeneralizedEnergyEquipartition2014}. This model predicts a ballistic regime at small times followed by a long time saturation and could be thus used to fit the experimental results. However, as obtained recently by Villalobos et al.~\cite{Villalobos2024}, the harmonic potential approximation to render the spherical confinement in our setup only holds when the particle's movements are limited to a small region near the equilibrium position and the normalized variance remains small ($\langle \tilde{\rho}^2 \rangle = \langle \rho^2 \rangle / R^2 \lesssim 0.04$). This limitation arises since the linear restoring force of the harmonic potential is only an approximation of the gravitational force acting on the particle near equilibrium. The approximation would work as long as the particle remains in contact with the outer drop surface, which is generally not the case (\textit{SI Appendix}, SI text and Fig. S2). Therefore, to correctly extract the active bath properties, we have developed a full 3D numerical model and a specific methodology to fit the experimental data to the model, as exposed in the following section.

\subsection*{Numerical solution of a 3D stochastic model for the probe kinematics}
We model the inner particle as a non-Brownian passive spherical particle of radius $R_i$ immersed in an active fluid under the influence of a gravitational force and a viscous drag arising from the active fluid in the low Reynolds number regime. Both the particle and the active fluid are confined within a spherical domain of outer radius $R_o$. The sedimentation speed of the particle due to the gravitational force is given by $v_s = \Delta M g/ \gamma$, where $\gamma$ is the effective viscous drag coefficient.

The equation of motion for the inner particle is described by the following overdamped equation:
\begin{equation}
\label{eq.Langevin}
\dot{\mathbf{r}} = \mathbf{u}(t) - v_s \hat{\mathbf{z}},
\end{equation}
where $\mathbf{r}$ is the position vector of the particle center and $\mathbf{u}(t)$ is the velocity field imposed by the bacteria. The spherical confinement is captured using a reflective boundary condition on the particle's position, given by $x^2 + y^2 + z^2 \leq R^2$~\cite{Volpe2014}.

We model the active fluid velocity $\mathbf{u}(t)$ as an Ornstein-Uhlenbeck process (OUP), which is a Gaussian stochastic process with a finite correlation time. The OUP is characterized by two parameters: the characteristic speed $u_b$ and the persistence time $\tau_b$. The correlation function of the OUP is given by $\langle u_i(t) u_j(t') \rangle = u_b^2 \delta_{ij} e^{-|t-t'|/\tau_b}$, where $i, j \in {x,y,z}$. The OUP satisfies the following stochastic differential equation:
\begin{equation}
\dot{\mathbf{u}} = -\frac{\mathbf{u}}{\tau_b} + \boldsymbol{\xi},
\end{equation}
where $\boldsymbol{\xi}$ is a white Gaussian noise with zero mean and correlation $\langle \xi_i(t) \xi_j(t') \rangle = 2 u_b^2 \tau_b^{-1} \delta_{ij} \delta(t-t')$.

To non-dimensionalize the equations, we introduce the following dimensionless parameters:
\begin{align}
\label{eq.adim_numbers}
\tilde{\ell}_b &= \frac{u_b \tau_b}{R}, \\
\tilde{\tau}_b &= \frac{v_s \tau_b}{R}.
\end{align}
Here, $\tilde{\ell}_b$ compares the persistence length of the bacterial bath, $u_b \tau_b$, to the confinement $R$, while $\tilde{\tau}_b$ compares the persistence time of the bath, $\tau_b$ to the sedimentation time of the particle, $\tau_s = R/v_s$. The dimensionless active bath speed is given by $\tilde{u}_b = \tilde{\ell}_b / \tilde{\tau}_b = u_b/v_s$. The resulting dimensionless equation of motion for the particle is (dimensionless variables are denoted by tildes):
\begin{equation}
\label{eq.Motion_dimless}
\frac{d \mathbf{\tilde r}}{d \tilde t} = \mathbf{\tilde u} - \hat{\mathbf z}.
\end{equation}
The dimensionless bacterial bath velocity satisfies the Langevin equation for the OUP:
\begin{equation}
\label{eq.AOUP}
\frac{d \mathbf{\tilde u}}{d \tilde t} = - \frac{\mathbf{\tilde u}}{\tilde \tau_b} + \boldsymbol{\tilde \xi} \left( \tilde t \right).
\end{equation}
Eqs.~(\ref{eq.Motion_dimless}) and (\ref{eq.AOUP}) are solved numerically with a forward Euler scheme. Simulations are performed for a wide range of dimensionless parameters $\tilde{\ell}_b$ and $\tilde{\tau}_b$, covering the experimentally relevant values.

\subsection*{Bath parameters extraction}
To extract the active bath parameters $\tau_b$ and $u_b$, for each experimental realization, the experimental MSD curves are compared with the simulated MSDs generated from the model described by Eqs.~(\ref{eq.Motion_dimless}) and (\ref{eq.AOUP}). To compare the experimental planar MSD curves to the non-dimensional numerical ones, the confinement $R$ and the sedimentation timescale $\tau_s$ are used to give adequate length and time scales. The confinement $R$ is directly obtained from independent experimental measurements, which is not the case for the sedimentation speed. The reason is that the drag coefficient $\gamma$ given by the Stokes law is strictly holding as a constant for an infinite homogeneous medium. Due to the nearby confining surface, one expects a more complex effective drag force and, in particular, an enhancement due to lubrication-like forces that develop for particles approaching significantly close to the droplet surface~\cite{rallison1978note, happel1983low}. Additionally, the viscosity $\eta$ of the bacterial suspension is expected to vary with bacteria concentration and thus affect the sedimentation time~\cite{lopez2015turning}. To account for these effects, the sedimentation time $\tau_s$ is left as a fitting parameter, together with $\tilde \tau_b$ and $\tilde \ell_b$, to best reproduce the experimental observations by minimizing the difference between the experimental and simulated MSDs (see Methods). The dimensionless bath parameters $\tilde{\ell}_b$ and $\tilde{\tau}_b$ obtained from the fit span approximately two decades each (\textit{SI Appendix}, Fig. S3). Figure~\ref{fig:fig_2} shows examples of the fitted MSD curves. The agreement between the experimental and simulated MSDs indicates that the model accurately captures the probe particle's dynamics.

\begin{figure}[h!]
    \centering
    \includegraphics[width=\linewidth]{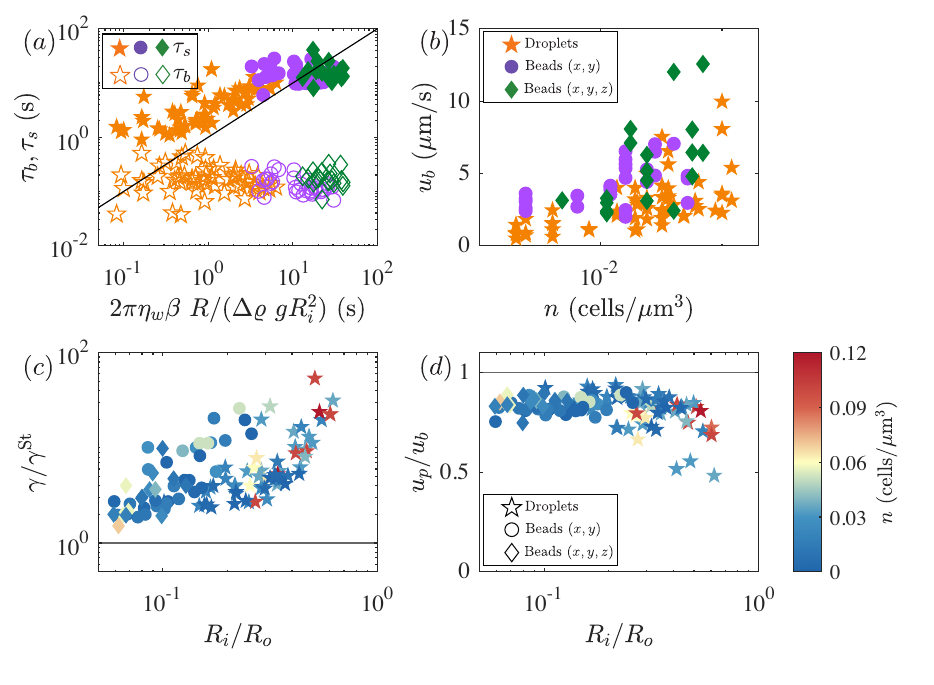}
     \caption{(a) Times obtained by the fitting protocol for the bath ($\tau_b$, open symbols) and sedimentation ($\tau_s$, solid symbols) as a function of the $\tau_s^\mathrm{St} = 2 \beta \pi \eta_w R / (\Delta \varrho g R_i^2)$. The solid line indicates the prediction of the sedimentation time with the Stokes drag coefficient, $\tau_s^\mathrm{St}$.
    (b) Bath velocity $u_b$, obtained by fitting the experimental MSD curves, as a function of the bacterial density $n$.
    (c) Experimentally obtained drag coefficient $\gamma$ normalized by the Stokes drag coefficient, $\gamma^\mathrm{St}$ as a function of the confinement $R_i/R_o$.
    (d) Particle speed $u_p$ normalized by the bath speed $u_b$. For reference, in (c) and (d), the horizontal solid line represents a ratio equal to one. The color bar is common for (c) and (d).}
    \label{fig:fig_3}
\end{figure}

From the fitting procedure, two timescales are extracted, $\tau_b$ and $\tau_s$, which are shown in Fig.~\ref{fig:fig_3}(a).
If lubrication-like forces and active stresses were neglected, the sedimentation time would be given by $\tau_s^\mathrm{St} = \gamma^\mathrm{St} R / (\Delta \varrho g R_i^3)$, where $\gamma^\mathrm{St} = 2 \beta \pi \eta_w R_i$ is the Stokes drag coefficient for the inner particle, with $\beta_\text{bead} = 3$ for the solid beads and $\beta_\text{drop} = (2 + 3 \eta_o/\eta_w) / (1 + \eta_o/\eta_w) \approx 1.1$ for double emulsions, being $\eta_w$ and $\eta_o$ the water and oil viscosity, respectively. For comparison, the extracted times are presented against $2 \beta \pi \eta_w R / (\Delta \varrho R_i^2)$. A separation between the two timescales is evident, with $\tau_s$ (solid symbols) having a scaling similar to $\tau_s^\mathrm{St}$, marked with a black line. For all experimental conditions, the sedimentation time is the largest, growing approximately linearly with $\beta R/(\Delta \varrho R_i^2)$ although with a higher value than $\tau_s^\mathrm{St}$. The memory time $\tau_b$ (open symbols), on the contrary, is rather constant and takes values between \SI{0.1}{\second} and \SI{0.6}{\second}, approximately, with no clear dependence on bacterial density $n$ nor on other control parameters. These values are compatible with previous measurements of tracer kinematics in bacterial suspensions~\cite{Kasyap2014, patteson2016particle, Lagarde_2020} and of the movement of drops driven by enclosed bacteria~\cite{Ramos2020}.

The bath velocity $u_b$, depicted in Fig.~\ref{fig:fig_3}(b), varies between approximately \SI{0.5}{\micro\meter/\second} and \SI{12}{\micro\meter/\second}. It displays a significant scattering but also an apparent tendency to increase with the bacterial density $n$. As a reference, the characteristic bacterial density for the onset of collective motion is $n \approx \SI{0.1}{cells/\micro\meter^3}$~\cite{patteson2016particle}. We do not observe any change of behavior at this concentration, except for a possible increase in data dispersion.

\subsection*{Friction coefficient}
The fact that $\tau_s > \tau_s^\mathrm{St}$ points on an expected increase of the effective friction coefficient $\gamma = \Delta M g \tau_s / R$ with respect to a mere Stokes prediction. The assessment of an effective friction coefficient $\gamma$ allows the extraction of bath parameters without any assumption on the bath viscosity. Figure~\ref{fig:fig_3}(c) shows the values of $\gamma$, normalized by the Stokes drag coefficient as a function of the confinement $R_i/R_o$. Notably, we do not observe a dependence of $\gamma$ on the bacteria concentration. This result, together with an increase of $\gamma$ with $R_i/R_o$, suggests that the friction enhancement can be viewed as an effective lubrication effect arising when the tracer particles are in close proximity to the confining walls. Note that this friction coefficient is an effective average value over a full trajectory. This lubrication increase may also stem from the transport of momentum from the inner fluid to the surrounding viscous oil via the fluid interface~\cite{Vincenti2019}. These data seem to split into two groups featuring a larger effective friction for the solid tracers when compared to the droplet tracers, probably because of the non-slip boundary condition at the surface of solid beads.

\subsection*{Particle velocities}
From the experimental MSD curves at short time scales, we can directly extract the mean persistence speed of the particle $u_p$ using the relation $\langle \Delta \bm \rho^2 (t) \rangle = u_p^2 t^2$. Figure~\ref{fig:fig_3}(d) shows the comparison between the particle speed $u_p$ and the bath speed $u_b$. Interestingly, the particle speed is always smaller than the bath velocity. The ratio $u_p/u_b$ is almost constant around $0.8$ with eventually, a slight decrease at higher confinement ratio $R_i/R_o$. For particles confined in a harmonic potential in an OUP bath, it has been shown that the particle speeds are always smaller than the bath as an effect of the bath memory, $u_p^2/u_b^2 = 1/(1 + \tau_b/\tau_s)$, which only equals one for a memoryless (Brownian) bath~\cite{maggiGeneralizedEnergyEquipartition2014, Villalobos2024}. The ratio of the two times is in the range $\tau_b/\tau_s \in [0.0035,0.56]$ [see Fig.~\ref{fig:fig_3}(a)], implying that for the harmonic model, the velocity ratio should be in the range $u_p/u_b\in[0.80, 0.998]$. The harmonic model, which, as discussed before, is not expected to be quantitatively valid for most of the experiments, underestimates the particle bath reduction as the experimental values lay in the range $u_p/u_b \in [0.48, 0.94]$.

\subsection*{Active bath diffusivity}
\begin{figure}[t!]
    \centering
    \includegraphics[width=\linewidth]{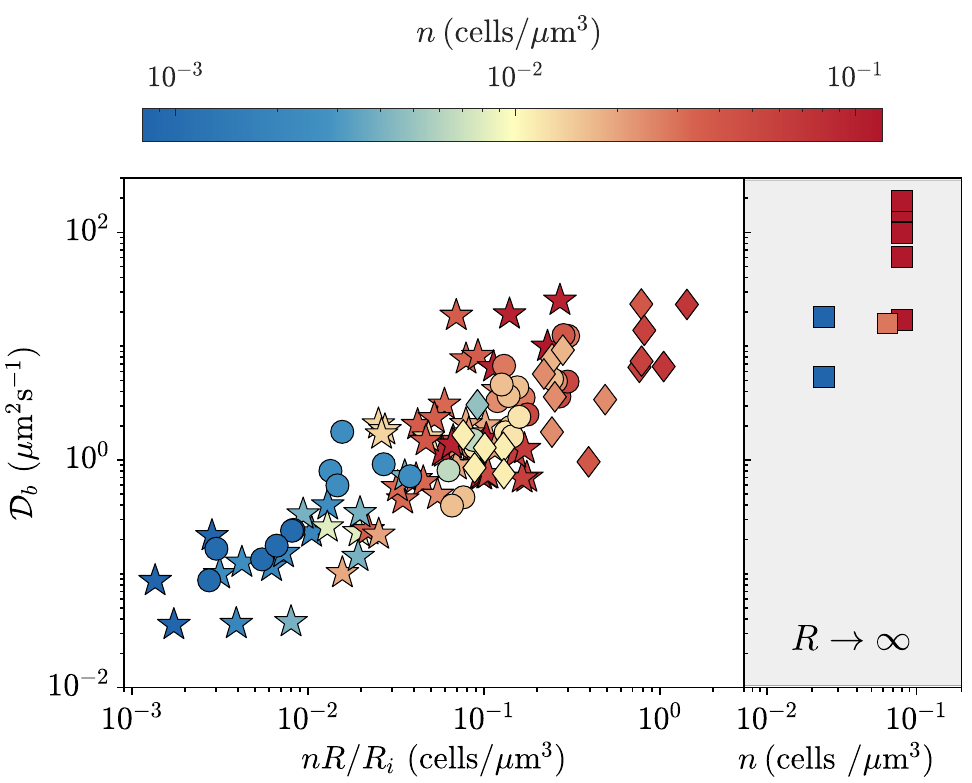}
    \caption{Bath diffusivity $\mathcal{D}_b = u_b^2 \tau_b$ as a function of $n R/R_i$. Symbols represent different experimental setups according to the legend in Fig.~\ref{fig:fig_3}(b). The square symbols in the shaded region denote the bath diffusivity for melamine beads with a radius of $R_i = \SI{2}{\micro\meter}$ in a suspension of bacteria within a flat chamber for densities in the range $n \in [\num{2e-2}, \num{8e-2}]~\si{cells/\micro\meter^3}$, serving as a reference for confinement approaching $R \to \infty$. $\mathcal{D}_b$ was determined in that case by fitting the MSD to $\langle \Delta \rho^2 \rangle = 2u_b^2 \tau_b^2 \left( t/\tau - 1 + e^{-t/\tau_b} \right)$.}
    \label{fig:fig_4}
\end{figure}

In spite of the fact that the bath velocities present an increasing trend with density, the measured data are highly scattered. However, the bath diffusivity defined as $\mathcal{D}_b = u_b^2 \tau_b$, shows an interesting collapse for all the data over three decades when plotted against $n R/R_i$, as shown in Fig.~\ref{fig:fig_4}. When $\mathcal{D}_b$ is plotted only against the density or the confinement, such collapse is not obtained, with data highly scattered (\textit{SI Appendix}, Fig. S4). This result shows that the bath activity, characterized by $\mathcal{D}_b$, is controlled not only by the bacterial density but also by the confinement, quantified as the available space relative to the particle size.

The increasing trend of $\mathcal{D}_b$ with $n R/R_i$ cannot follow for infinitely large values, and the presented results must correspond to a small value expansion in $n R/R_i$. For comparison, in Fig.~\ref{fig:fig_4}, we present the bath diffusivity obtained using melamine beads of radius $R_i = \SI{2}{\micro\meter}$ that sediment in a large flat chamber filled with a bacterial suspension, which serves as a reference for a case with outer confining radius going to infinity. In this configuration, there is no saturation in the MSD as the beads can reach the diffusive regime, allowing direct extraction of the diffusivity (\textit{SI Appendix}, SI Text). The bath diffusivity in this case is larger than the one obtained in droplets, consistent with the previous discussion that the collapse is valid for small values of $n R/R_i$. Note that in the $R \to \infty$ limit, the bath diffusivity shows a monotonic increasing dependence with density, implying that the collapse is not uniform for large values of $n R/R_i$ and new dependencies should be incorporated.

With these results, the confined active bath is well described as an Ornstein-Uhlenbeck process with correlation
\begin{align} \label{eq.finalbath}
\langle u_i(t) u_j(t') \rangle = \frac{\mathcal{D}_b}{\tau_b} \delta_{ij} e^{-|t-t'|/\tau_b},
\end{align}
with $\tau_b$ roughly constant and $\mathcal{D}_b$ increasing with bacterial density and the available space.

\section*{Conclusions}
By integrating experiments, simulations, and theoretical modeling, we characterized the active bath properties of motile bacterial suspensions confined within a spherical domain. Buoyant tracers immersed in the suspension were tracked across a wide range of bacterial densities, varying both the tracer and confining radii. Our findings reveal that the active bath can be effectively described as a stochastic driving flow on the tracer, with velocity fluctuations following an Ornstein-Uhlenbeck process [\eqref{eq.finalbath}]. The stochastic model is governed by two key parameters: memory time and diffusivity. The bath memory, $\tau_b$ remains approximately constant, showing no dependence on bacterial density $n$, tracer radius $R_i$, or confining radius $R_o$. In contrast, the bath diffusivity $\mathcal{D}_b$ increases with bacterial density and, notably, with the available space $R \equiv R_o-R_i$. This interplay between confinement and diffusivity underscores how the capacity of an active bath to mix and transport particles is strongly influenced by the spatial constraints. Our results corroborate previous studies suggesting that bacterial bath properties depend on bacterial concentration~\cite{wuParticleDiffusionQuasiTwoDimensional2000, Mino2011}, confinement~\cite{Mino2013, yeActiveNoiseExperienced2020}, and tracer size~\cite{Kasyap2014, patteson2016particle}. By synthesizing these elements, we demonstrate a data collapse spanning three orders of magnitude when diffusivity is plotted as a function of  $n R/R_i$.

The finding that spatial confinement shapes bath activity may appear counterintuitive. However, while boundary conditions do not influence the properties of fluids at thermal equilibrium, they are known to fundamentally alter the dynamics and characteristics of non-equilibrium systems~\cite{Galajda2007, DiLeonardo2010, Sokolov2010, Berke2008}. To fully disentangle the effects of boundaries from the intrinsic properties of active baths, future research should focus on rigorously elucidating the role of confinement in the dynamics of non-dilute bacterial baths.

\matmethods{\section*{Methods}

\subsection{Bacterial suspensions}
Wild-type \textit{Escherichia coli} (strain W3110) was used in the experiments. A frozen stock stored at -20°C was diluted in 10 mL of lysogeny broth (LB) and incubated overnight in a shaking incubator at 30°C and 210 rpm. The overnight culture was then diluted 100-fold in fresh LB and allowed to grow for an additional 3.5 hours to mid-exponential phase, reaching an optical density at 600 nm of OD$_{600} = 0.6 \pm 0.1$.

To prepare the bacterial suspension for the experiments, the culture was centrifuged at $2100 g$ for \SI{10}{\minute}. The supernatant was removed, and the bacterial pellet was resuspended in a minimal motility buffer composed of \SI{10}{\milli\mol/\liter} potassium phosphate, \SI{0.1}{\milli\mol/\liter} EDTA, and \SI{10}{\milli\mol/\liter} sodium lactate. In this minimal medium, bacteria can swim but do not divide, maintaining a constant cell concentration throughout the experiments. The final concentration of bacteria in the suspension was varied between OD$_{600} = 1$ and OD$_{600} = 150$, which corresponds to cell number density between \SI{8e-4}{cell/\micro\meter^3} and \SI{1.2e-1}{cell/\micro\meter^3}.

\subsection{Preparation of double emulsion drops}
Double emulsions were generated by mechanically mixing the prepared bacterial suspension with a mixture of hexadecane oil (Sigma-Aldrich, H6703) and Span 80 surfactant (Sigma-Aldrich, S6760). The oil phase was prepared by dissolving \SI{2}{\percent} Span 80 by weight in hexadecane. \SI{10}{\micro\liter} of the bacterial suspension was added to \SI{1}{\milli\liter} of the oil phase, and the mixture was pipetted to form aqueous droplets dispersed in the oil. Approximately \SI{10}{\percent} of the resulting droplets were double emulsions, each containing a single oil droplet within the aqueous phase.

To facilitate the observation and recording of isolated double emulsion droplets, the mixture was further diluted 100-fold in the hexadecane-surfactant solution. This dilution step reduced the overall droplet concentration, enabling the imaging of individual droplets without interference from neighboring ones.

\subsection{Preparation of drops with bead tracers}
Particle-loaded droplets were prepared using the same procedure as for the double emulsions, with the addition of passive melamine particles of density $\varrho_p = \SI{1.51}{\gram/\centi\meter^3}$ (Microparticles GmbH) to the bacterial suspension prior to mixing with the oil phase. Two different particle sizes were used: \SI{4}{\micro\meter} and \SI{8}{\micro\meter} in diameter. The concentration of particles in the suspension was adjusted such that the majority of the resulting droplets contained either one or no particles, with a small fraction containing up to three particles per droplet. For the experiments, only droplets containing a single particle were selected for analysis.

\subsection{Observation chamber}
The observation chambers are squares of inner side $L = \SI{1}{\centi\meter}$ and height $h = \SI{400}{\micro\meter}$ fabricated in SU-8 photoresist (GM1075, Gersteltec Sarl) by optical lithography on a $\SI{50.8}{\milli\meter}$ diameter circular glass wafer. A volume of approximately \SI{300}{\micro\liter} of hexadecane-bacterial emulsion was deposited on the chamber and closed with a coverslip ensuring that no bubbles were present in the sample. Prior to assembly, the glass surfaces were thoroughly cleaned with ethanol. To ensure that the bacterial droplets maintained a spherical shape when reaching the bottom of the observation chamber, the glass surfaces were rendered hydrophobic using Aquapel.

\subsection{Correction of detection noise}
Tracking inner droplets and beads in the $(x,y)$ plane was performed using a custom-made Matlab algorithm based on the Hough transform. Due to the highly fluctuating background behind the particles, the detection algorithm gives noisy trajectories. These fluctuations between consecutive frames give rise to an artificial subdiffusive behavior at short times in the mean squared displacement. Smoothing the raw trajectory data is not a solution for this detection error, as we found that the estimated bath parameters are very sensitive to the smoothing kernel applied to the trajectories (\textit{SI Appendix}, SI Text and Fig. S5). Instead, we treat the detection error as follows.

Let $x_i$ be the actual $x$ coordinate of the detected object at discrete times steps $t_i = i\Delta t$, with $i = 0,1,2,\dots$ (the analysis is analogous for the $y$ coordinate). Assuming that the coordinate measurement process has a normally distributed noise with zero mean, $\sigma_i$, which is independent of the particle position, the measured position is $x'_i = x_i + \sigma_i$. Defining the noise autocorrelation function  $g_{i-k} = \langle \sigma_i \sigma_k \rangle$, with $g_i$ an even function of $i$, the measured MSD is given by
\begin{align}
    \langle \Delta x'^2_i \rangle &=\langle (x'_{i+k} - x'_k)^2 \rangle_k,\\
    &= \langle\left[(x_{i+k} - x_k) + (\sigma_{i+k} - \sigma_{k})\right]^2\rangle_k,\\
    &= \langle \Delta x_i^2 \rangle + \langle \sigma^2_{i+k} + \sigma_k^2 - 2\sigma_{i+k}\sigma_k \rangle_k,\\
   \langle\Delta x'^2_i\rangle  &=\langle \Delta x_i^2\rangle + 2(g_0-g_i).
    \end{align}
Considering that the detection noise has short correlation time, $\langle \sigma_i \sigma_k \rangle = \sigma^2 \delta_{ik}$, the measured MSD is
\begin{equation}
   \langle\Delta x'^2_i\rangle = \langle \Delta x_i^2\rangle + \begin{cases}
   0 \quad &i=0,\\
   2\sigma^2 \quad &i>0.
   \end{cases}
\end{equation}

Assuming that the MSD of the probe is ballistic at short times, $\langle \Delta x_i^2 \rangle = (u_p t_i)^2$, we can obtain $\sigma^2$ by fitting the short-time measured MSD as $\langle\Delta x'^2_i\rangle = 2\sigma^2 + (u_p t_i)^2$ for $i>0$. Then, the corrected MSD, $\langle \Delta x_i^2\rangle$, is obtained by subtracting the offset $2\sigma^2$ from the measured one.

\subsection{Fitting of bath parameters and sedimentation timescale}
The fitting procedure scans the simulation MSD curves performed in the $(\tilde \tau_b, \tilde \ell_b)$ space for the one that most closely resembles each experimental curve. The dimensionless numerical MSDs are scaled using the experimental length scale $R$ in each case. Dimensionless times of the simulations are scaled using the sedimentation time $\tau_s$. However, since the experimental sedimentation time is not available, it is left as a third fitting parameter. The error function to minimize is then
\begin{equation}
\chi^2(\tilde \tau_b, \tilde \ell_b, \tau_s) = \sum_{i} \frac{(\log(\mathrm{MSD}_\mathrm{exp}(t_i)) - R^2 \log(\mathrm{MSD}_\mathrm{sim}(\tau_s \tilde t_i)))^2}{t_i}.
\end{equation}
The evaluation times $t_i$ are the experimental time points and the simulation curves are interpolated in those points for the evaluation of $\chi^2$. Normalizing the sum by $t_i$ emulates, for equally spaced time points, a sum over logarithmically spaced time points. With this, together with the use of the logarithm of the MSDs in the error function, we give equal weight to both early and late time behaviors, ensuring that the fitting procedure captures the dynamics across different time scales.

We compare each experimental dataset with a large set of simulations (approximately 40,000) covering a wide range of dimensionless parameters $\tilde{\tau}_b \in [10^{-4}, 10^2]$ and $\tilde{\ell}_b \in [10^{-4}, 10]$. For each pair $(\tilde \tau_b, \tilde \ell_b)$, the error function is minimized with respect to $\tau_s$. The optimized set of parameters $(\tilde \tau_b, \tilde \ell_b, \tau_s)$ corresponds to the one that yields the global minimum of $\chi^2$.
}

\showmatmethods{}

\newpage

\acknow{This work was partially funded by ANID through Fondecyt grants No. 1210634 and 1220536, and Millennium Science Initiative Program NCN19\_170. E.C and A.L. acknowledge the ``PushPull'' (ANR-22-CE30-0038) grant. Z.L, A.L, E.C and T.L-L acknowledge the MITI-``Auto-Organisation 2021'' for support. Z.L. was funded by a special CNRS-INSIS post-doctoral Grant. We thank Olivier Dauchot, from Gulliver Lab, UMR CNRS 7083, for the use of the “clinofocalscope”. CVC was funded by ANID Doctoral grant No 21201766. The authors acknowledge the support of the Franco-Chilean program Programa de Cooperación Científica ECOS-ANID ECOS210012 / ECOS-Sud C21E05.}

\showacknow{}

\bibsplit[26]
\bibliography{Active_Bath_Droplet_V3}

\end{document}